\documentclass[twocolumn,showpacs,tightenlines,preprintnumbers,prl,amsmath,amssymb]{revtex4}

\usepackage{graphicx}
\usepackage{epsfig}
\usepackage{dcolumn}
\usepackage{bm}

\def\he3{$^3$He}
\def\al{\alpha}

\def\ga{\gamma}

\def\w{\omega}
\def\higs{HI$\vec{\gamma}$S}
\def\bra{\langle}
\def\ket{\rangle}

\def\an{\alpha^{(n)}}
\def\bn{\beta^{(n)}}

\newcommand{\veps}{{\hat\epsilon}}
\newcommand{\vepsprime}{{\hat\epsilon\, '}}

\newcommand{\mpi}{m _\pi}

\begin{document}

\title{Investigating Neutron Polarizabilities through Compton Scattering on \he3}

\author{Deepshikha Choudhury$^1,^2$}\email{choudhur@phy.ohiou.edu}
\author{Andreas Nogga$^3$}
\author{Daniel R. Phillips$^1$}
\affiliation{$^1$Department of Physics and Astronomy, Ohio
University, Athens, OH 45701, $^2$Department of Physics, George
Washington University, Washington DC 20052, $^3$Institut
f$\ddot{u}$r Kernphysik, Forschungszentrum J$\ddot{u}$lich,
J$\ddot{u}$lich, Germany}

\date{\today}

\begin{abstract}
We examine manifestations of neutron electromagnetic polarizabilities
in coherent Compton scattering from the Helium-3 nucleus. We calculate
$\gamma$\he3~elastic scattering observables using chiral perturbation
theory to next-to-leading order (${\mathcal O}(e^2 Q)$). We find that
the unpolarized differential cross section can be used to measure
neutron electric and magnetic polarizabilities, while two
double-polarization observables are
sensitive to different linear combinations of the four neutron spin
polarizabilities.
\end{abstract}

\maketitle

The theory that describes the internal dynamics of the neutron is
quantum chromodynamics (QCD). The neutron has zero charge, but
higher electromagnetic moments encode the strong-interaction
dynamics which governs its structure. These quantities therefore
provide tests of our understanding of QCD.  For example, an early
success of the $SU(3)$ quark picture was its prediction of magnetic
moments, $\vec{\mu}$, for the neutron and other strongly-interacting
particles (hadrons). Magnetic moments are a first-order response to
an applied magnetic field. In this paper we will be concerned with
{\it electromagnetic polarizabilities}, which quantify the
second-order response of a particle to electromagnetic fields. The
two most basic polarizabilities are the electric and magnetic ones,
$\alpha$ and $\beta$, which measure the ability of an applied
electric or magnetic field to produce an induced dipole moment. The
Hamiltonian for a neutral particle in applied electric and magnetic
fields, $\vec{E}$ and $\vec{B}$, is then:
\begin{equation}
H=- \vec{\mu}\cdot \vec{B}-2\pi\left[{\alpha}\,\vec{E}^2+
{\beta}\,\vec{B}^2\right], \label{eq:H1}
\end{equation}
where we have worked up to second order in $\vec{E}$ and $\vec{B}$,
and have not yet considered terms which involve derivatives of these
fields. For a spin-half particle consideration of first-order
derivatives allows four new structures which are second order in
$\vec{E}$ and $\vec{B}$~\cite{barry}. They are:
\begin{eqnarray}
&&-2\pi \left[ (-\ga_1 - \ga_3)\,\vec{\sigma}\cdot\vec{E}\times\dot{\vec{E}}+ \ga_4\,\vec{\sigma}\cdot\vec{B}\times\dot{\vec{B}} -(\ga_2+\ga_4)\right. \nonumber\\
&& \quad \left. \sigma_i\,(\nabla_iE_j+\nabla_jE_i)\,B_j+
\,\ga_3\,\sigma_i\,(\nabla_iB_j+\nabla_jB_i)\,E_j\right]
\label{eq:H2}
\end{eqnarray}
The coefficients $\gamma_1$--$\gamma_4$ are the ``spin
polarizabilities''. This paper will argue that for the neutron, the
most basic and stable neutral hadron, $\alpha$, $\beta$, and
$\gamma_1$--$\gamma_4$ can be extracted from Compton scattering on
\he3.

Polarizabilities such as those in Eqs.~(\ref{eq:H1}) and
(\ref{eq:H2}) can be accessed in Compton scattering because
the Hamiltonian (\ref{eq:H1}) yields an
amplitude for Compton scattering from a neutron
target of the form:
\begin{equation}
T_{\ga n}= \sum \limits_{i=1 \ldots 6} A_i^{(n)} (\w,\, \theta) t_i.
 \label{eq:amp}
\end{equation}
Here $t_1$--$t_6$ are invariants constructed out of the photon
momenta and polarization vectors ($\veps$ and $\vepsprime$), and, in
the case of $t_3$--$t_6$, the neutron spin, e.g. $t_1=\vepsprime
\cdot \veps$ and $t_3=i \vec{\sigma} \cdot (\vepsprime \times
\veps)$. The $A_i$'s are Compton structure functions. The $\omega^2$
terms of $A_1$ and $A_2$ involve $\alpha$ and
$\beta$~\cite{comment}, while the $\omega^3$ terms of $A_3$--$A_6$
depend on $\gamma_1$--$\gamma_4$ in various combinations.

For the proton, an expression similar to Eq.~(\ref{eq:amp}) but
supplemented by the Thomson term, $-\frac{e^2}{M} \vepsprime \cdot
\veps$, applies. The larger cross sections that result from the
addition of this term lend themselves to low-energy measurements
from which $\alpha^{(p)}$ and $\beta^{(p)}$ can be extracted. A
considerable number of $\ga$p experiments over the past decade had
this as their goal~\cite{Schumacher06}. A combined analysis of their
low-energy differential cross section (DCS) data
yields~\cite{silas2}:
\begin{eqnarray}
\alpha^{(p)}&=& (12.1 \pm 1.1({\rm stat.}))^{+0.5}_{-0.5}({\rm th.}) \times 10^{-4} \, {\rm fm}^3, \label{eq:pexp1} \\
\beta^{(p)}&=& (3.4 \pm 1.1({\rm stat.}))^{+0.1}_{-0.1}({\rm th.})
\times 10^{-4} \, {\rm fm}^3. \label{eq:pexp2}
\end{eqnarray}

No elastic Compton scattering measurement of the $\gamma_i^{(p)}$'s
has yet been performed, but they affect double-polarization
observables. Of these, $\Delta_z$ and $\Delta_x$ are defined by
taking the beam helicity to be along $\hat{z}$; then $\Delta_z$
($\Delta_x$) is the difference between the DCS when the target is
spin-polarized along $+\hat{z}$ ($+\hat{x}$) and along $-\hat{z}$
($-\hat{x}$). For $\omega < m_\pi$ the $\ga_i^{(p)}$s affect
$\Delta_z$ and $\Delta_x$ because of interference between $A_3^{(p)}
\ldots A_6^{(p)}$ and $A_1^{(p)}$ in the expressions for these
observables~\cite{bkmrev}. An experiment which exploits this
interference to probe $\gamma_1^{(p)}$--$\gamma_4^{(p)}$ has been
proposed for the High-Intensity Gamma-ray Source (HI$\vec{\gamma}$S)
at TUNL~\cite{gap}.

However, neither polarized nor unpolarized Compton scattering
experiments can be directly performed on the neutron, since it is
not a stable target. A variety of techniques have been proposed to
extract $\alpha^{(n)}$ and $\beta^{(n)}$, including neutron
scattering from the Coulomb field of ${}^{208}$Pb and Compton
scattering on the deuteron---both elastic and quasi-free. The most
accurate numbers come from the last technique and yield (in units of
$10^{-4} \, {\rm fm}^3$)~\cite{kossert}:
\begin{equation}
\an-\bn=(9.8 \pm 3.6 {\rm (stat)} \pm 2.2 {\rm (mod.)}
{}^{+2.1}_{-1.1} {\rm (sys)}). \label{eq:nexp}
\end{equation}

These numbers represent a fascinating interplay of long-distance ($r
\sim 1/m_\pi$) and short-distance ($r \sim 1/\Lambda$) dynamics. The
dominant piece of $\an$ is due to the ``cloud'' of virtual pions
that surrounds the neutron. But there are also significant
contributions from short-distance physics---especially in $\bn$.
This interplay can be systematically computed in baryon chiral
perturbation theory ($\chi$PT), a low-energy effective theory that
encodes the low-energy symmetries of QCD and the pattern of their
breaking (see Ref.~\cite{bkmrev} for a review). Observables in
$\chi$PT are computed in an expansion in powers of $Q \equiv
\frac{p,m_\pi}{\Lambda}$, where $\Lambda$ is the excitation energy
of the lightest state not explicitly included in the theory. At
${\mathcal O}(e^2 Q)$ there are no contributions to the $\gamma$n
amplitude from a short-distance $\gamma$n operator. The prediction
for the $\gamma$n amplitude comes from nucleon-pole, pion-pole, and
one-pion-loop diagrams, with the latter capturing the dominant piece
of the ``pion cloud". This ${\mathcal O}(e^2 Q)$ calculation yields
the entire dependence of $A_1^{(n)}$--$A_6^{(n)}$ on photon energy
and scattering angle up to corrections of ${\mathcal
O}(\frac{\omega}{\Lambda})$. The ${\mathcal O}(\omega^2)$ and
${\mathcal O}(\omega^3)$ non-pole pieces of $A_1^{(n)}$--$A_6^{(n)}$
then give~\cite{bkmrev, ulf}:
\begin{eqnarray}
  \alpha^{(n)}&=&10 \beta^{(n)}=\frac{5 e^2 g_A^2}{384 \pi^2 f_\pi^2 m_\pi}=12.2
  \times 10^{-4}~{\rm fm}^3;\label{eq:alphabeta}\\
\gamma_1^{(n)}&=&2 \gamma_2^{(n)}= 4 \gamma_3^{(n)}=-
4\gamma_4^{(n)}= 4.4 \times 10^{-4}~{\rm fm}^4.\label{eq:gammas}
\end{eqnarray}
(The $\gamma_i^{(n)}$'s can also be written in terms of $g_A$,
$f_\pi$, and $m_\pi$.) The contributions of short-distance physics to
Eq.~(\ref{eq:alphabeta}) are suppressed by one power of $Q$, and to
Eq.~(\ref{eq:gammas}) are suppressed by two powers of $Q$.
In addition, $\chi$PT
predicts that $\alpha^{(p)}$, $\beta^{(p)}$, and the $\gamma_i^{(p)}$'s
 are the same as the corresponding neutron quantities---at this
order. These ${\mathcal O}(e^2 Q)$ predictions of $\chi$PT agree
with the numbers in Eqs.~(\ref{eq:pexp1}) and (\ref{eq:nexp}) within
the experimental error bars.

We now examine how the predictions of Eqs.~(\ref{eq:alphabeta}) and
(\ref{eq:gammas}) can be tested in elastic $\ga$\he3~scattering. The
scattering amplitude is written as
\begin{equation}
{\mathcal M}=\bra \Psi_f|{\hat O}|\Psi_i \ket , \label{calM}
\end{equation}
with $|\Psi_i\ket$ and $|\Psi_f\ket$ being the anti-symmetrized
\he3~ wavefunctions. The results quoted in this letter have been
calculated using a wavefunction obtained from the Idaho-N$^3$LO
chiral potential~\cite{entem} together with the NNLO chiral 3N
force~\cite{andreas3nf}. For reviews of $\chi$PT applied to nuclear
forces see Ref.~\cite{cptrev}. Note, however, that aspects of this
power-counting are still under discussion~\cite{pc}.

The operator ${\hat O}$ in Eq.~(\ref{calM}) is the irreducible
amplitude for elastic scattering of real photons from the NNN
system, calculated in $\chi$PT up to ${\mathcal O}(e^2 Q)$. This is
next-to-leading order (NLO), a lower order than was used to obtain
$|\Psi \ket$, and so our calculation is chirally consistent only to
NLO. At NLO $\hat{O}$ has a one-body part
\begin{equation}
{\hat O}^{1B}={\hat O}^{1B}(1)+{\hat O}^{1B}(2)+{\hat O}^{1B}(3),
\label{ch6-eq2}
\end{equation}
with ${\hat O}^{1B}(a)$ being the $\gamma$N amplitude where the
external photon interacts with nucleon `$a$'. ${\hat O}^{1B}(a)$
(supplemented by what turn out to be very small corrections for the
boost from the $\gamma$N c.m. frame to the $\gamma$NNN c.m. frame)
follows from Eq.~(\ref{eq:amp}) and can be found in
Refs.~\cite{bkmrev, silas1}. Meanwhile the two-body part of
$\hat{O}$ is:
\begin{equation}
{\hat O}^{2B}={\hat O}^{2B}(1,2)+{\hat O}^{2B}(2,3)+{\hat
O}^{2B}(3,1), \label{ch6-eq3}
\end{equation}
and it represents a sum of two-body mechanisms where the external
photons interact with the pair `$(a,b)$'.  At ${\mathcal O}(e^2 Q)$
this operator encodes the physics of two photons coupling to a
single pion exchange inside the \he3~nucleus.  (We do not have to
include any irreducible three-body Compton mechanisms in our
calculation because they appear at the earliest at ${\mathcal O}(e^2
Q^3)$.) We use the expression for $\hat{O}^{2B}$ given in
Ref.~\cite{silas1}. This incorporates the few-nucleon physics that
corresponds to the pion-cloud dynamics which yields
Eqs.~(\ref{eq:alphabeta}) and (\ref{eq:gammas}). As such it must be
included on an equal footing with the polarizability effects that
are our focus.  The resulting $\hat{O}^{2B}$ gives a significant
contribution to the $\gamma$d cross section, and is an important
piece of the $\chi$PT calculations that provide a good description
of the extant $\gamma$d DCS
data~\cite{silas1,silas2,robertdelta,robert3}. We now simplify
Eq.~(\ref{calM}) to:
\begin{equation}
{\mathcal M} = 3\bra \Psi_f|\frac{1}{2} \big( {\hat O}^{1B}(1)+{\hat
O}^{1B}(2) \big)+{\hat O}^{2B}(1,2)|\Psi_i\ket , \label{calM2}
\end{equation}
using the Faddeev decomposition of $|\Psi\ket$. The structure of the
calculation is then similar for the one- and two-body parts. We
calculate ${\cal M}$ on a partial-wave Jacobi basis. Convergence of
the results with respect to the angular-momentum expansion was
confirmed. For details on the calculational procedure see
Ref.~\cite{long}.
\begin{figure}[htbp]
\epsfig{figure=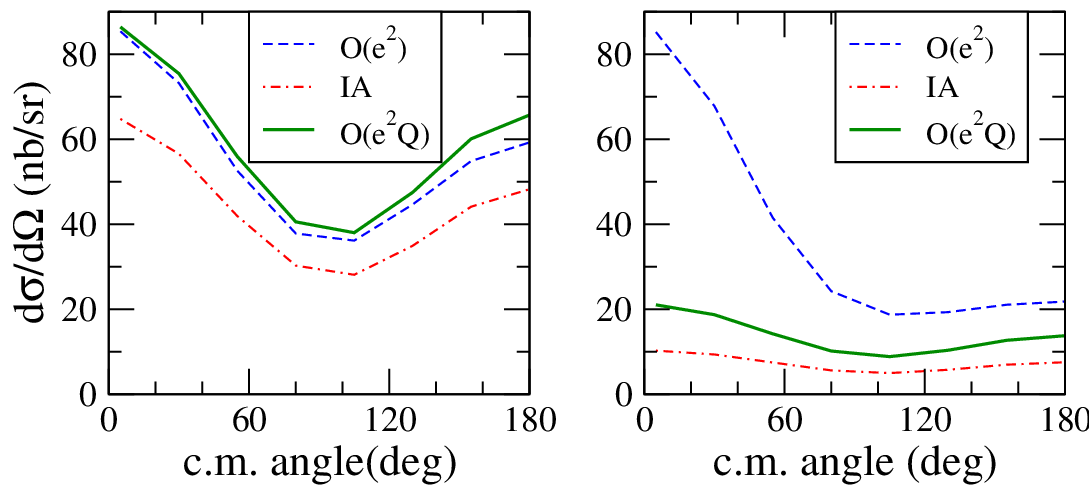, height=1.4in}
\caption{\label{fig0} Comparison of different c.m.-frame DCS
calculations at 60~MeV (left panel) and 120~MeV (right panel).}
\end{figure}

The amplitude~(\ref{calM2}) is now used to calculate observables. In
Fig.~\ref{fig0} we plot our ${\mathcal O}(e^2 Q)$ $\chi$PT DCS
predictions for coherent $\ga$\he3~scattering. The two panels are
for $\w=$ 60 and 120 MeV. Both show three different DCS
calculations---${\mathcal O}(e^2)$, $IA$ (Impulse Approximation) and
${\mathcal O}(e^2 Q)$. The ${\mathcal O}(e^2)$ calculation includes
only the proton Thomson term, since that is the $\gamma$N amplitude
in $\chi$PT at that order. The $IA$ calculation is done up to
${\mathcal O}(e^2 Q)$ but does not have any two-body contribution.
As expected, we see that there is a sizeable difference between the
$IA$ and the ${\mathcal O}(e^2 Q)$ DCS: the two-body currents are
important and cannot be neglected. Also, we see that the difference
between ${\mathcal O}(e^2)$ and ${\mathcal O}(e^2 Q)$ is very small
at 60 MeV---showing that $\chi$PT may converge well there---and
gradually increases with energy.  This is partly because the
fractional effect of $\alpha^{(n)}$ and $\beta^{(n)}$ increases with
$\w$.
\begin{figure}[htbp]
\epsfig{figure=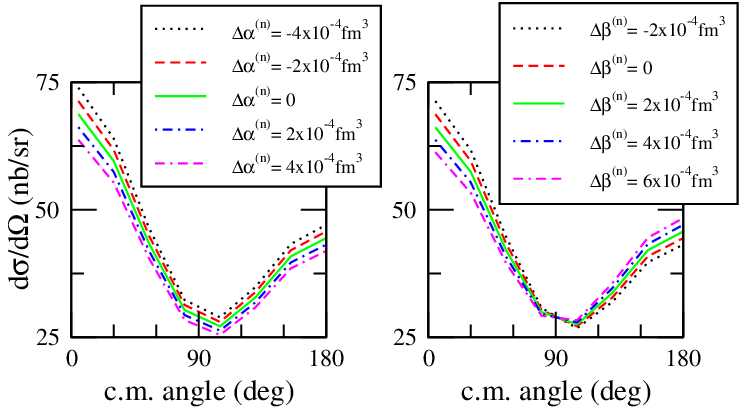, height=1.8in}
\caption{\label{fig1} The c.m.-frame ${\mathcal O}(e^2 Q)$ DCS at
80~MeV with varying $\Delta \an$ (left panel) and $\Delta \bn$
(right panel).}
\end{figure}

To quantify this, in Fig.~\ref{fig1} we plot the ${\mathcal O}(e^2
Q)$ DCS at 80~MeV obtained when we add shifts, $\Delta \al^{(n)}$
and $\Delta \beta^{(n)}$, to the ${\mathcal O}(e^2 Q)$ values of the
neutron electric and magnetic polarizabilities (\ref{eq:alphabeta}).
We take $\Delta \an $ in the range $(-4 \ldots 4) \times
10^{-4}$~fm$^3$ and $\Delta \bn $ between $(-2 \ldots 6) \times
10^{-4}$~fm$^3$. This allows us to assess the impact that one set of
higher-order mechanisms has on our ${\mathcal O}(e^2 Q)$
predictions. Two features of Fig.~\ref{fig1} are particularly
notable. First, sensitivity to ${\bn}$ vanishes at $\theta =
90^{\circ}$ because $\al^{(n)}$ and $\beta^{(n)}$ enter $A_1^{(n)}$
in the combination $\al^{(n)}+\beta^{(n)}\cos\theta$. Thus, ${\an}$
and ${\bn}$ can be extracted independently from the same experiment.
Second, the absolute size of the shift in the DCS due to $\Delta
\an$ and $\Delta \bn$ is roughly the same for all energies. This
suggests that measurements could be done at $\w \approx 80$ MeV,
where the count rate is higher, and the contribution of higher-order
terms in the chiral expansion should be smaller.

We have estimated the uncertainty due to short-distance physics in
the three-nucleon system by using a variety of \he3~wave functions
generated using various NN interactions with and without a
corresponding 3N force. This produced changes of $\lesssim 15\%$ in
the DCS at 120~MeV.

Before examining double-polarization observables in $\gamma$\he3~
scattering we try to develop some intuition for the $\gamma$\he3~
amplitude. Since \he3~is a spin-${1\over2}$ target the matrix
element~(\ref{calM2}) can be decomposed in the same fashion as was
the neutron's Compton matrix element in Eq.~(\ref{eq:amp}).
\begin{equation}
T_{\ga ^3He}= \sum \limits_{i=1 \ldots 6} A_i^{^3He}(\w,\theta) t_i;
\;\;\; A_i^{^3He} = A_i^{1B}+A_i^{2B},
 \label{aihe3}
\end{equation}
where $A_i^{1B}$ ($A_i^{2B}$) comes from considering the matrix
element of the one-body (two-body) operators in Eq.~(\ref{calM2}),
and the structures $t_3$--$t_6$ now involve the nuclear---not the
neutron---spin. However, in \he3~the two proton spins are---to a
good approximation---anti-aligned, so the nuclear spin is largely
carried by the unpaired neutron~\cite{he3pol}. We find that the
${\mathcal O}(e^2 Q)$ two-body currents $A_1^{2B}$ and $A_2^{2B}$
are numerically sizeable, but $A_3^{2B}$--$A_6^{2B}$ are negligible.
Hence, to the extent that polarized \he3~is an effective neutron, we
expect $A_i^{{}^3He}=A_i^{(n)}$ for $i=3$--$6$. Using
Eq.~(\ref{aihe3}) to translate this into predictions for $\Delta_z$
and $\Delta_x$ shows that the effects of
$\gamma_1^{(n)}$--$\gamma_4^{(n)}$ will be enhanced  in these
observables by interference with $A_1^{{}^3He}$. But $A_1^{^3He}$
is---at least at $\w \approx 80$ MeV---dominated by the contribution
of the two protons, and so we anticipate a more marked signal from
the neutron spin polarizabilities than is predicted for the
corresponding $\ga$d observables~\cite{me}.
\begin{figure*}[bhtp]
\epsfig{figure=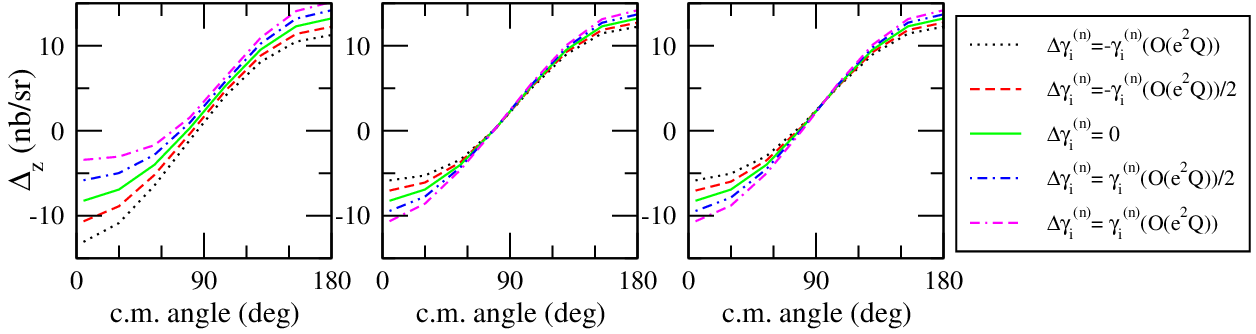, height=1.4in}
\caption{\label{fig2} $\Delta_z$ at $\omega=$120~MeV with
(left-to-right) $\ga_1^{(n)}$, $\ga_2^{(n)}$, and $\ga_4^{(n)}$
varied one at a time. For ${\mathcal O}(e^2 Q)$ $\ga_i^{(n)}$'s see
Eq.~(\ref{eq:gammas}).}
\end{figure*}
\begin{figure}[htbp]
\epsfig{figure=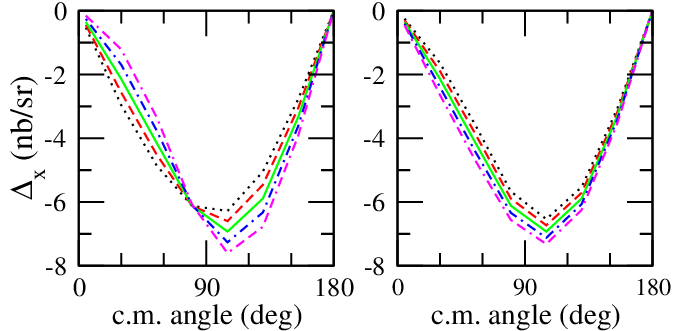, height=1.4in}
\caption{\label{fig3} $\Delta_x$ (c.m. frame) at $\omega=$120 MeV
when $\ga_1^{(n)}$ (left) and $ \ga_4^{(n)}$ (right) are varied one
at a time. Legend as in Fig.~\ref{fig2}.}
\end{figure}

We emphasize that these arguments are meant only as a guide to the
physics of our exact ${\mathcal O}(e^2 Q)$ calculation. Our \he3~
wave function is obtained by solving the Faddeev equations with NN
and 3N potentials derived from $\chi$PT. All of the effects due to
neutron depolarization and the spin-dependent pieces of
$\hat{O}^{2B}$ are included in our calculation of the amplitude
(\ref{calM}). This yields the results for $\Delta_z$ and $\Delta_x$
shown in Figs.~\ref{fig2} and \ref{fig3}. There we have proceeded
analogously to our computations of the $\gamma$\he3~DCS, this time
varying the neutron spin polarizabilities and seeing the effect on
$\Delta_z$ and $\Delta_x$. Fig.~\ref{fig2} indicates that $\Delta_z$
is quite sensitive to $\ga_1^{(n)}$, $\ga_2^{(n)}$, and
$\ga_4^{(n)}$. With the expected photon flux at an upgraded
HI$\vec{\gamma}$S such effects can be measured~\cite{gao}. If this
can be done as a function of $\theta$ we can extract the combination
$\ga_1^{(n)} - (\ga_2^{(n)} + 2\ga_4^{(n)})\cos \theta$.  Turning to
$\Delta_x$, Fig.~\ref{fig3} shows that varying $\ga_1^{(n)}$ or
$\ga_4^{(n)}$ produces appreciable effects in $\Delta_x$---but in a
different combination to the sensitivity in $\Delta_z$.  Use of
different \he3~wave functions alters these predictions for
$\Delta_x$ and $\Delta_z$ by $\lesssim 7.5 \%$. For a more detailed
discussion see~\cite{long}.  Thus, $\Delta_z$ and $\Delta_x$ are
sensitive to two different linear combinations of $\ga_1^{(n)}$,
$\ga_2^{(n)}$, and $\ga_4^{(n)}$ and their measurement should
provide an unambiguous extraction of $\gamma_1^{(n)}$, as well as
constraints on $\gamma_2^{(n)}$ and $\gamma_4^{(n)}$.

These $\ga$\he3~scattering calculations are the first calculations
for this reaction, and there is significant scope for improvement.
Computing of the NNLO (${\mathcal O}(e^2 Q^2)$) pieces of the
$\gamma$NNN operator $\hat{O}$ would allow a more detailed
assessment of the pattern of convergence. When this is done we
anticipate three kinematic domains where convergence may be slow.
First, since we use the heavy-baryon formulation of $\chi$PT, the
pion-production threshold is at $\omega=m_\pi$, rather than in the
correct position for $\ga$\he3~scattering which is $\sim$4~MeV
lower. An estimate of the impact of this discrepancy on observables
suggests a $\sim 5\%$ difference in the DCS, and $\lesssim 3\%$ in
$\Delta_z$ and $\Delta_x$, at 100 MeV. Second, the power counting we
used is not valid at energies $\lesssim \frac{\mpi^2}{M}$. For
instance, it does not reproduce the correct $\omega=0$ (Thomson)
limit for the nuclear target, since the terms in $\hat{O}$ that
restore that limit are higher-order effects when $\omega \sim
m_\pi$. (A recent computation for $\gamma$d scattering verifies that
they are indeed small for $\omega \geq 80$ MeV~\cite{robert3}.) We
therefore expect that assessment of these two classes of
corrections, while an important check on our results, will not
significantly alter them. We believe that the most important
correction will come from the inclusion of $\Delta$(1232) degree of
freedom. $\ga$d scattering calculations which included such effects
found a sizeable impact on the DCS at backward angles for $\omega
\approx 100$~MeV~\cite{robertdelta}.

Our results for $\ga$\he3~scattering are obtained from $\chi$PT NN
\& NNN interactions and $\ga$N \& $\ga$NN operators, and are
accurate to NLO in the chiral expansion. These first results on this
reaction suggest that $\an$ and $\bn$ can be extracted from the
$\ga$$^3$He DCS and compared to results from $\ga$d experiments at
MAXLab~\cite{fissum}. Meanwhile, two different linear combinations
of $\ga_1^{(n)}$, $\ga_2^{(n)}$, and $\ga_4^{(n)}$ can be
constrained by measurements of double-polarization observables at
facilities such as \higs. This would provide new information on
neutron polarizabilities.

We thank H.~Gao, A.~Nathan, and L.~Platter for useful conversations.
This work was carried out under grant DE-FG02-93ER40756 of the
US-DOE and NSF grant PHY-0645498 (DC). The wave functions have been
computed at the NIC, J\"ulich.


\begin{thebibliography}{99}
\bibitem{comment}
These `Compton' polarizabilities equal the `static' ones
(Eq.~(\ref{eq:H1})) only in the non-relativistic limit, but the
$\sim 5\%$ difference is understood and calculable.
\end{thebibliography}
\end{document}